%% file: Letter.tex
\documentclass[10pt,prl,reprint,showpacs,%
               amssymb,amsmath,%
               floatfix,linenumbers,%
               preprintnumbers,%
               footinbib,longbibliography]{revtex4-1}
\usepackage{atlasphysics}
\usepackage{graphicx}
\usepackage{multirow}
\usepackage[hang,center]{subfigure}
\usepackage{hyperref}
\begin{document}
\nolinenumbers
\title{Measurement of Dijet Azimuthal Decorrelations in $pp$
  Collisions at $\sqrt{s}=7$~TeV} 
\author{The ATLAS Collaboration} 
\preprint{CERN-PH-EP-2011-014; Submitted to PRL}
\begin{abstract}

\noindent Azimuthal decorrelations between the two central jets with the largest transverse momenta are sensitive to the dynamics of events with multiple jets. We present a measurement of the normalized differential cross section based on the full dataset ($\totalluminoerr$) acquired by the ATLAS detector during the 2010 $\rts = 7\tev$ proton-proton run of the LHC. The measured distributions include jets with transverse momenta up to $1.3\tev$, probing perturbative QCD in a high energy regime.
\end{abstract}
\maketitle

The production of events containing high transverse-momentum (\pT) jets is a key signature of quantum chromodynamic (QCD) interactions
between partons in $pp$ collisions at large center-of-mass energies (\rts). The Large Hadron Collider (LHC) opens a window into the dynamics of interactions with high-\pT\ jets in a new energy regime of $\rts = 7\tev$. QCD predicts the decorrelation in the azimuthal angle between the two most energetic jets, \dphi, as a function of the number of partons produced. Events with only two high-\pT\ jets have small azimuthal decorrelations, $\dphi\sim\pi$, while $\dphi<\!\!<\pi$ is evidence of events with several high-\pT\ jets. QCD also describes the evolution of the shape of the \dphi\ distribution, which narrows with increasing leading jet \pT.  Distributions in \dphi\ therefore test perturbative QCD (pQCD) calculations for multiple jet production without requiring the measurement of additional jets. Furthermore, a detailed understanding of events with large azimuthal decorrelations is important to searches for new physical phenomena with dijet signatures, such as supersymmetric extensions to the Standard Model~\cite{randallsusy}. 

In this Letter, we present a measurement of dijet azimuthal decorrelations with jet $\pT$ up to $1.3\tev$ as measured by the ATLAS detector, beyond the reach of previous colliders. The normalized differential cross section \xsec\ is based upon an integrated luminosity \totallumi~\cite{lumi}. The \dphi\ distribution is normalized by the inclusive dijet cross
section, $\sigma$, integrated over the same phase space. This construction minimizes experimental and theoretical uncertainties. Previous measurements of \dphi\ from the D0~\cite{d0dphi} and CMS~\cite{cmsdphi} collaborations are extended here to higher jet \pT\ values. 

Jets are reconstructed using the \AKT\ algorithm~\cite{antikt} (implemented with \fastjet~\cite{fastjet}) with radius $R = 0.6$, and the jet four-momenta are constructed from a sum over its constituents, treating each as an $(E,\vec{p})$ four-vector with zero mass.  The
\AKT\ algorithm is well-motivated since it is infrared-safe to all
orders, produces geometrically well-defined cone-like jets, and
is used for pQCD calculations (from partons), event
generators (from stable particles), and the detector (from energy
clusters~\cite{inclusivejet}). The azimuthal decorrelation, \dphi,
is defined as the absolute value of the difference in azimuthal angle
between the jet with the highest \pT\ in each event, \ptone, and the jet with the second-highest
\pT\ in the event. There are nine analysis regions in \ptone, where the lowest region is bounded by $\ptone > 110\gev$ and the highest region requires $\ptone > 800\gev$~\cite{inclusivejet}.  Only jets with $\pt > \minpt$ and $|y| < \etaaccept$, where $y$ is the jet rapidity~\cite{coordinatesystem}, are considered. The two leading jets that define $\dphi$ are required to satisfy $|y| < \etacut$, restricting the measurement to a central $y$ region where the momentum fractions $(x)$ of the interacting partons are roughly equal and the experimental acceptance for multijet production is increased. In this region where $0.02 \lesssim x \lesssim 0.14$, the parton distribution function (PDF) uncertainties are typically $\pm3\%$ (at fixed factorization scale)~\cite{mstw}. The cross sections, measured over the range $\pi/2\leq\dphi\leq\pi$ and normalized independently for each analysis region, are compared with expectations from a pQCD calculation~\cite{nlojet} that is next-to-leading order (NLO) in three-parton production. The perturbative prediction for the cross section is $\mathcal{O}(\alpha_s^4)$, where $\alpha_s$ is the strong coupling constant.

The angular decorrelation is sensitive to multijet configurations such as those
produced by event generators like \sherpa~\cite{sherpa}, which matches higher-order tree-level pQCD diagrams with a dipole parton-shower model~\cite{dipoleshower}. Samples for $2\rightarrow2-6$ jet production are combined using an improved CKKW matching scheme~\cite{ckkw}. The progression of the parton shower is vetoed to avoid double counting of emissions. Event generators such as \pythia~\cite{pythia} and \herwig~\cite{herwig} use $2\rightarrow 2$ leading order pQCD matrix elements matched with phenomenological parton-cascade models to simulate higher-order QCD effects. Such models have been successful at reproducing other QCD processes measured by the ATLAS collaboration~\cite{inclusivejet,wjet}.
  
The ATLAS detector~\cite{atlasdetector,CSC}
consists of an inner tracking system surrounded by a thin
superconducting solenoid providing a 2T magnetic field,
electromagnetic and hadronic calorimeters, and a muon spectrometer
based on large superconducting toroids. Jet measurements depend
most heavily on the calorimeters.  The electromagnetic calorimeter is
a lead liquid-argon (LAr) detector with an accordion geometry.  Hadron calorimetry is based on
two different detector technologies, with scintillator tiles or LAr as
the active medium, and with either steel, copper, or tungsten as the
absorber material. The pseudorapidity ($\eta$)~\cite{coordinatesystem} and $\phi$
segmentations of the calorimeters are sufficiently fine to ensure that
angular resolution uncertainties are negligible compared to other
sources of systematic uncertainty.  

\begin{figure}[t!] 
\begin{center}
\includegraphics[width=1.0\columnwidth,trim=0 30 0 10]{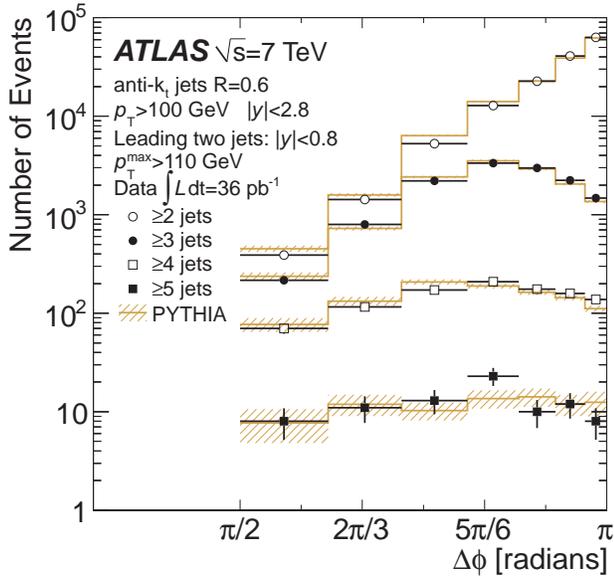}
\caption{\label{fig:dphi_reco} The \dphi\ distribution for
$\geq2$, $\geq3$, $\geq4$, and $\geq5$ jets
with $\pT > 100\gev$.  Overlaid on the calibrated but otherwise uncorrected data (points) are results from \pythia\ processed through the detector simulation (lines). All uncertainties are statistical only. }
\end{center}
\end{figure}
\begin{figure}[t!] 
\begin{center}
{\includegraphics[width=1.0\columnwidth,trim=12 55 18 25]{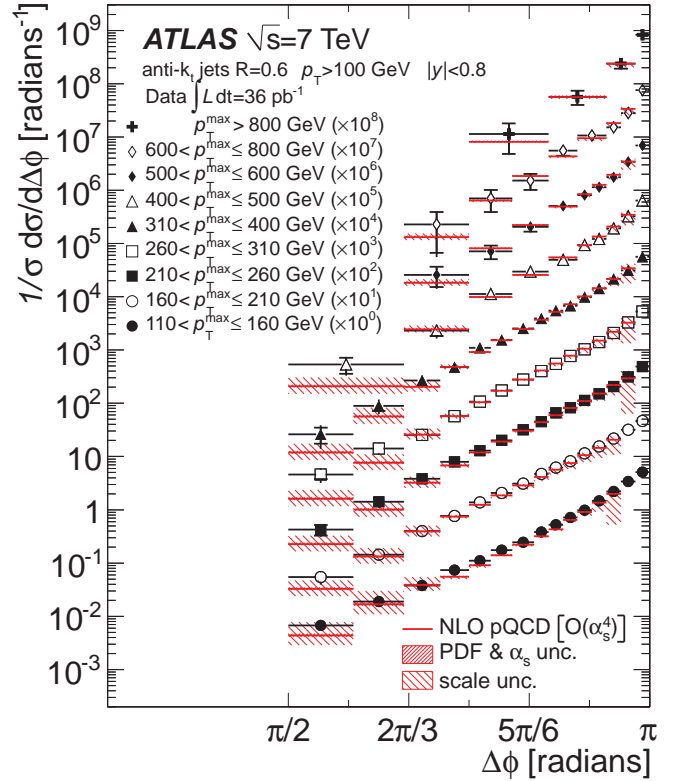}}%
\caption{\label{fg:crosssection} The differential cross section
  \xsec\ binned in nine \ptone\ regions.  Overlaid on the
  data (points) are results from the NLO pQCD calculation. The error bars on the data points indicate the statistical (inner error bar) and systematic uncertainties added in quadrature in this and subsequent figures. The theory
  uncertainties are indicated by the hatched regions. Different bins in \ptone\ are scaled by multiplicative factors of
  ten for display purposes. The region near the divergence at $\dphi\to\pi$
  is excluded from the calculation.
}
\end{center}
\end{figure}

A hardware-based calorimeter jet trigger identified events of interest; the decision was further refined
in software~\cite{atlasdetector,CSC}.  Events with at least one jet
that satisfied a minimum transverse energy (\ET) requirement
were recorded for further analysis. The events in each \ptone\ range are selected
by a single trigger with a given \ET\ threshold, and the lower end of the range is chosen above the jet \pT\ at which that trigger is $\approx100\%$ efficient. Three sets of triggered events with different integrated luminosity are considered: $2.3\,\ipb$ for $110 < \ptone \leq 160\gev$, $9.6\,\ipb$ for $160 < \ptone \leq 260\gev$, and $36\,\ipb$ for $\ptone > 260\gev$~\cite{lumi}.  Events are also required to have a reconstructed primary vertex within $15\,$cm in $z$ of the center of the detector; each vertex had $\geq5$ associated tracks. The inputs to the \AKT\ jet algorithm are clusters of calorimeter cells seeded by cells with energy that is significantly above the measured noise~\cite{inclusivejet}. Jets reconstructed in the detector,
whether in data or the GEANT4-based simulation~\cite{atlassimulation,geant4}, are corrected for the effects of
hadronic shower response and detector-material distributions using a
\pT- and \eta-dependent calibration~\cite{inclusivejet} based on the detector simulation and validated with
extensive test-beam~\cite{atlasdetector} and collision
data~\cite{EOPNote} studies.  Jets likely to have arisen from detector noise or cosmic
rays are rejected~\cite{JetQualityNote}.

The resulting \dphi\ distribution is shown in Fig.~\ref{fig:dphi_reco} for jets with $\pT > 100\gev$. There are $146788$ events in the data sample, $85$ of which have at least five jets with $\pT > 100\gev$. Also shown is the \pythia\ sample with MRST 2007 $\rm{LO}^*$ PDF~\cite{LO*} and ATLAS MC09 underlying event tune~\cite{MC09tune}, processed through the full detector simulation and normalized to the number of events in the data sample. Two- and three-jet production primarily populates the region $2\pi/3 <\dphi < \pi$ while smaller values of \dphi\ require additional activity such as soft radiation or more jets in an event. Fig.~\ref{fig:dphi_reco} illustrates that the decorrelation increases when a third high-\pT\ jet is also required. Events with additional high-\pT\ jets widen the overall distribution.

The measured differential \dphi\ distributions in data are corrected in a single step with a bin-by-bin unfolding method~\cite{inclusivejet} to compensate for trigger and detector inefficiencies and the effects of finite experimental resolutions. These correction factors, evaluated using the \pythia\ sample, lie within $\pm9\%$ of unity. The leading sources of systematic uncertainty on the normalized cross section are the jet energy scale calibration ($2 - 17\%$)~\cite{inclusivejet}, the bin-by-bin unfolding method ($1 - 19\%$), and the jet energy and position resolutions ($0.5 - 5\%$). The ranges in parentheses represent the magnitude of the uncertainties near $\pi$ and $\pi/2$, respectively, and correspond to the analysis region with the smallest statistical uncertainty ($160 < \ptone \leq 210\gev$). Uncertainties due to multiple $pp$ interactions in the same beam crossing ($<0.8\%$ on the cross section for all analysis regions) are included in the evaluation of the jet energy scale uncertainties.

\begin{figure}[t!] 
\begin{center}
{\includegraphics[width=1.0\columnwidth,trim=10 0 4 0]{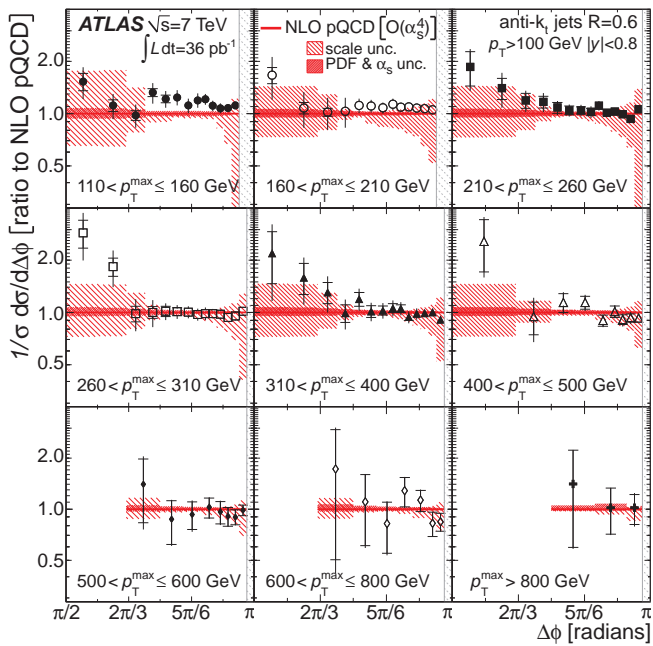}}%
\caption{\label{fg:crosssectionratioNLO} Ratio of the differential
  cross section \xsec\ measured in data with respect to expectations
  from NLO pQCD (points).  The theory uncertainties are indicated by the hatched regions. The region near the divergence at $\dphi\to\pi$ is excluded from the comparison.}
\end{center}
\end{figure}

\begin{figure}[t!] 
\begin{center}
{\includegraphics[width=1.0\columnwidth,trim=10 0 4 0]{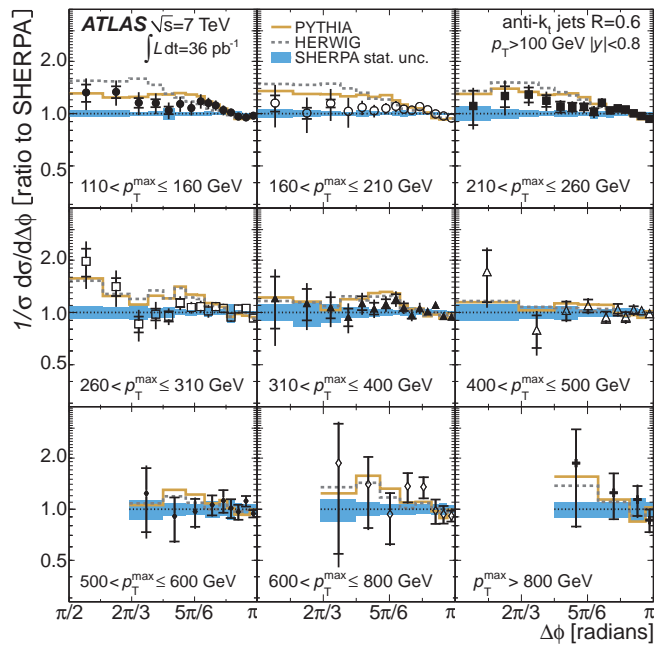}}%
\caption{\label{fg:crosssectionratio} Ratio of the differential cross
  section \xsec\ measured in data with respect to the result from
  \sherpa\ (points). The shaded region indicates the \sherpa\ statistical
  uncertainty.  Predictions from \pythia\ and \herwig, also in ratio
  to \sherpa, are displayed as lines.}
\end{center}
\end{figure}
  
The normalized differential cross section is shown for each of the nine \ptone\ analysis regions as a function of \dphi\ in Fig.~\ref{fg:crosssection}. As \ptone\ increases, and the probability for the emission of a hard third jet is reduced, the fraction of events near $\pi$ becomes larger. Overlaid on the data are the results from a NLO pQCD $[\mathcal{O}(\alpha_s^4)]$ calculation, NLOJET++~\cite{nlojet} with fastNLO~\cite{fastnlo} and using the MSTW 2008 PDF~\cite{mstw}. The factorization and renormalization scales are set to \ptone\ and are varied independently up and down by a factor of two to determine the scale uncertainties. The scale uncertainties are larger between $\pi/2 < \dphi < 2\pi/3$ where the pQCD calculation is effectively leading order in four-parton production. The PDF uncertainties are treated as the envelope of the $68\%$ CL uncertainties from MSTW 2008~\cite{mstw}, NNPDF $2.0$~\cite{nnpdf}, and CTEQ 10~\cite{ct10}, and are combined with the uncertainties resulting from an $\alpha_s$ variation of $\pm0.004$; the $\alpha_s$ contributions dominate. The calculation is corrected for non-perturbative effects due to hadronization and the underlying event~\cite{mc10,rivet}; the correction is smaller than $3\%$. The fixed-order calculation fails near $\dphi \to \pi$ where soft processes dominate and contributions from logarithmic terms are enhanced. Figure~\ref{fg:crosssectionratioNLO} displays the ratio of the cross section with respect to the NLO calculation. In most regions, the theory is consistent with the data. However, the prediction in the range $110 < \ptone < 160\gev$ is relatively low in the central region of $\dphi$ where the scale uncertainties are small.

The data are also compared with predictions from \sherpa, \pythia, and \herwig\ in Fig.~\ref{fg:crosssectionratio}. The leading-logarithmic approximations used in these event generators' parton-shower models effectively regularize the divergence at $\dphi \to \pi$; all three provide a good description of the data in this region. In the region $\pi/2 < \dphi < 5\pi/6$, where multijet contributions are significant, this observable distinguishes between the three generators. \sherpa, which explicitly includes higher-order tree-level diagrams, performs well in most \dphi\ and \ptone\ regions. Having phenomenological parameters that have been adjusted to previous ATLAS measurements, \pythia~\cite{mc10} and \herwig~\cite{MC09tune} also describe the data.

In summary, we present a measurement of dijet azimuthal decorrelations in events produced in $pp$ collisions at $\rts=7\tev$. The normalized differential cross sections are based on the full dataset ($\totalluminoerr$) collected by the ATLAS collaboration during the 2010 run of the LHC. Expectations from NLO pQCD $[\mathcal{O}(\alpha_s^4)]$ and those of several event generators successfully describe the general characteristics of our measurements, including the increasing slope of the \dphi\ distribution with \ptone\ and the shape near $\dphi\sim\pi/2$ where events with multiple jets make a considerable contribution. Our data, which include jets with $\pT$ values that significantly exceed earlier measurements, explore QCD in a new kinematic region.

\nolinenumbers
\begin{acknowledgments}

We wish to thank CERN for the efficient commissioning
and operation of the LHC during this initial high-energy data-taking
period as well as the support staff from our institutions without whom
ATLAS could not be operated efficiently.

We acknowledge the support of ANPCyT, Argentina; YerPhI, Armenia; ARC,
Australia; BMWF, Austria; ANAS, Azerbaijan; SSTC, Belarus; CNPq and
FAPESP, Brazil; NSERC, NRC and CFI, Canada; CERN; CONICYT, Chile; CAS,
MOST and NSFC, China; COLCIENCIAS, Colombia; MSMT CR, MPO CR and VSC
CR, Czech Republic; DNRF, DNSRC and Lundbeck Foundation, Denmark;
ARTEMIS, European Union; IN2P3-CNRS, CEA-DSM/IRFU, France; GNAS,
Georgia; BMBF, DFG, HGF, MPG and AvH Foundation, Germany; GSRT,
Greece; ISF, MINERVA, GIF, DIP and Benoziyo Center, Israel; INFN,
Italy; MEXT and JSPS, Japan; CNRST, Morocco; FOM and NWO, Netherlands;
RCN, Norway; MNiSW, Poland; GRICES and FCT, Portugal; MERYS (MECTS),
Romania; MES of Russia and ROSATOM, Russian Federation; JINR; MSTD,
Serbia; MSSR, Slovakia; ARRS and MVZT, Slovenia; DST/NRF, South
Africa; MICINN, Spain; SRC and Wallenberg Foundation, Sweden; SER,
SNSF and Cantons of Bern and Geneva, Switzerland; NSC, Taiwan; TAEK,
Turkey; STFC, the Royal Society and Leverhulme Trust, United Kingdom;
DOE and NSF, United States of America.

The crucial computing support from all WLCG partners is gratefully acknowledged, in particular from CERN and the ATLAS Tier-1 facilities at
TRIUMF (Canada), NDGF (Denmark, Norway, Sweden), CC-IN2P3 (France),
KIT/GridKA (Germany), INFN-CNAF (Italy), NL-T1 (Netherlands), PIC
(Spain), ASGC (Taiwan), RAL (UK) and BNL (USA) and in the Tier-2
facilities worldwide.

\vfill\eject

\end{acknowledgments}

\bibliography{Letter}

\onecolumngrid
\vfill
\input{atlas_authlist}

\end{document}

%% file: atlas_authlist.tex
\begin{flushleft}
{\Large The ATLAS Collaboration}

\bigskip

G.~Aad$^{\rm 48}$,
B.~Abbott$^{\rm 111}$,
J.~Abdallah$^{\rm 11}$,
A.A.~Abdelalim$^{\rm 49}$,
A.~Abdesselam$^{\rm 118}$,
O.~Abdinov$^{\rm 10}$,
B.~Abi$^{\rm 112}$,
M.~Abolins$^{\rm 88}$,
H.~Abramowicz$^{\rm 153}$,
H.~Abreu$^{\rm 115}$
% ...
\bigskip

% ...
$^{10}$ Institute of Physics, Azerbaijan Academy of Sciences, H. Javid Avenue 33, AZ 143 Baku, Azerbaijan\\
$^{11}$ Institut de F\'isica d'Altes Energies, IFAE, Edifici Cn, Universitat Aut\`onoma  de Barcelona,  ES - 08193 Bellaterra (Barcelona), Spain\\
% ...
$^{48}$ Albert-Ludwigs-Universit\"{a}t, Fakult\"{a}t f\"{u}r Mathematik und Physik, Hermann-Herder Str. 3, D - 79104 Freiburg i.Br., Germany\\
$^{49}$ Universit\'e de Gen\`eve, Section de Physique, 24 rue Ernest Ansermet, CH - 1211 Geneve 4, Switzerland\\
% ...
$^{88}$ Michigan State University, Department of Physics and Astronomy, High Energy Physics Group, East Lansing, MI 48824-2320, United States of America\\
% ...
$^{111}$ University of Oklahoma, Homer L. Dodge Department of Physics and Astronomy, 440 West Brooks, Room 100, Norman, OK 73019-0225, United States of America\\
$^{112}$ Oklahoma State University, Department of Physics, 145 Physical Sciences Building, Stillwater, OK 74078-3072, United States of America\\
% ...
$^{115}$ LAL, Univ. Paris-Sud, IN2P3/CNRS, Orsay, France\\
% ...
$^{118}$ Oxford University, Department of Physics, Denys Wilkinson Building, Keble Road, Oxford OX1 3RH, United Kingdom\\
% ...
$^{153}$ Tel Aviv University, Raymond and Beverly Sackler School of Physics and Astronomy, Ramat Aviv, IL - Tel Aviv 69978, Israel\\
% ...
\end{flushleft}


%% file: Letter.bbl
\begin{thebibliography}{29}
\bibitem [{coo()}]{coordinatesystem}%
  \BibitemOpen
  \href@noop {} {}\bibinfo {note} {{ATLAS uses a right-handed coordinate system
  with its origin at the nominal interaction point in the center of the
  detector. Cylindrical coordinates $(r,\phi)$ are used in the transverse
  plane, $\phi$ being the azimuthal angle around the beam axis. The
  pseudorapidity is defined in terms of the polar angle $\theta$ as
  $\eta=-\ln\tan(\theta/2)$. Rapidity is defined as
  $y=\frac{1}{2}\ln[(E+p_z)/(E-p_z)]$, where $E$ is the energy and $p_z$ is the
  longitudinal component of the momentum along the beam
  direction.}}\BibitemShut {Stop}%
\end{thebibliography}
